*Classification*:  Physical Sciences/Chemistry and Physics

*Title*:  **Voltage tuning of vibrational mode energies in single-molecule junctions**

*Authors*:  Yajing Li[1], Peter Doak[2,3], Leeor Kronik[4], Jeffrey B. Neaton[3,5,6], Douglas Natelson[1,7*]

[1]*Department of Physics and Astronomy,  MS 61, Rice University, Houston, TX 77005*

[2]*Department of Chemistry, University of California at Berkeley, Berkeley, CA 94720*

[3]*Molecular Foundry, Lawrence Berkeley National Laboratory, Berkeley, CA 94720*

[4]*Department of Materials and Interfaces, Weizmann Institute of Science, Rehovoth 76100, Israel*

[5]*Department of Physics, University of California at Berkeley, Berkeley, CA 94720*

[6]*Kavli Energy Nanosciences Institute at Berkeley, Berkeley, CA 94720*

[7]*Department of Electrical and Computer Engineering, MS 366, Rice University, Houston, TX 77005*

**Corresponding author**:
    Prof. Douglas Natelson
    Email:  natelson@rice.edu

*Abstract*:

**Vibrational modes of molecules are fundamental properties determined by intramolecular bonding, atomic masses, and molecular geometry, and often serve as important channels for dissipation in nanoscale processes. Although single-molecule junctions have been employed to manipulate electronic structure and related functional properties of molecules, electrical control of vibrational mode energies has remained elusive. Here we use simultaneous transport and surface-enhanced Raman spectroscopy measurements to demonstrate large, reversible, voltage-driven shifts of vibrational mode energies of $C_{60}$ molecules in gold junctions. $C_{60}$ mode energies are found to vary approximately quadratically with bias, but in a manner inconsistent with a simple vibrational Stark effect. Our theoretical model suggests instead that the mode shifts are a signature of bias-driven addition of electronic charge to the molecule. These results imply that voltage-controlled tuning of vibrational modes is a general phenomenon at metal-molecule interfaces and is a means of achieving significant shifts in vibrational energies relative to a pure Stark effect.**

*Significance statement*:  Like guitar strings, molecules have characteristic vibrational frequencies, set by the strength of chemical bonding between the atoms. In an experiment using a special antenna for light, we have found that applying an electrical voltage to a single buckyball molecule systematically lowers its vibrational frequencies, indicating that the bonds are weakened.  We can explain this observation in terms of a very simple, general model, in which the applied voltage slightly increases the amount of negative charge on the molecule, thus tuning the chemical bond strength.  This may be generally useful in understanding and controlling the mechanical properties of molecules.

Mechanical couplings between atoms within molecules, manifested through vibrational spectra, are critically important in many processes at the nanoscale, from energy dissipation to chemical reactions. These couplings originate from the self-consistent electronic structure and ionic positions within the molecule.(1) Vibrational spectroscopy examines this bonding, and advanced time-resolved techniques(2-5) can manipulate vibrational populations. Single-molecule junctions(6, 7) have also proven to be valuable tools for examining vibrational physics. Previous work has shown that vibrational frequencies can be altered in mechanical break junctions if the chemical linkage to the moving contacts is sufficient to strain bonds in the molecule,(8, 9) but has also shown vibrations to be unaffected when the linkage to the contacts is less robust(10). Controllably altering vibrational energies in the steady state is difficult, however. Electric fields can redistribute the molecular electron density and shift vibrational modes in the vibrational Stark effect(11), enabling spectroscopic probes of local static electric fields in charge double layers(12, 13) and bio systems(14-16). However, other physics may also be relevant, and studies of electrical tuning of molecular vibrational energies in single- or few-molecule-based solid-state junctions, which often provide clarity that is difficult to obtain from measurements of molecular ensembles, have been lacking.

Surface-enhanced Raman spectroscopy (SERS)(17, 18) in which surface plasmons enhance the Raman scattering rate for molecules, opens up the possibility of performing detailed vibrational studies at the single-molecule level. Plasmonic junctions between extended electrodes(19-25) show correlations of Raman response and conductance implying single or few molecule sensitivity, and enable studies of vibrational physics as a function of electrical bias. Spectral diffusion is often observed in single-molecule SERS experiments(18, 25, 26), and there is some preliminary evidence of bias-driven mode shifts in such junctions(23), with the mechanisms of these phenomena remaining unclear.

We report vibrational mode softening in $C_{60}$ molecules on the order of tens of wavenumbers, approximately quadratic in the external DC bias, $V$, applied across such a junction. We compare these observations with density functional theory (DFT) calculations to determine the underlying mechanism. The calculations suggest that the systematic softening, its magnitude, and its detailed functional dependence on $V$ are inconsistent with a pure vibrational Stark effect. Instead, changes in molecular charge with bias(27) result in vibrational shifts that closely resemble those observed in the experiments, both in magnitude and sign. This reveals a general physical mechanism, expected to have implications for other systems and measurements.

Figure 1 shows a typical Raman spectrum from a $C_{60}$-containing junction, prepared by electromigration(28) of a lithographically defined Au constriction on an oxidized Si substrate. This junction is a useful test system, because $C_{60}$ is known to adsorb sufficiently strongly to Au to allow the formation of reliable and reproducible junctions.(29-31) The fabrication and detailed measurement procedures are discussed in Methods. The incident wavelength for the Raman measurements is 785 nm, and the extended electrode design allows the application of a DC bias, $V$, across the junction and the flow of current through it. The sharp mode at 520 cm$^{-1}$ is from the Si substrate, and the modes between 1000 cm$^{-1}$ and 1600 cm$^{-1}$ are broadly consistent with expectations from previous $C_{60}$ single molecule SERS experiments and with our own calculations. We note that the close association of the molecule with the surface, necessary for SERS measurements, may result in chemical and symmetry changes that can turn previously Raman-inactive normal modes into active ones.(32) Because each of our devices produces a unique Au junction possessing a specific molecular association, we track all normal modes of the molecules in the calculations. For the isolated $C_{60}$ molecule, only the $A_g$ and $H_g$ modes are Raman active. They are labeled in Figures 4d, S4, S5, and S6.

As in previous studies[20, 21, 23, 25, 33], correlations as a function of time are observed between spectral intensity fluctuations ("blinking") and the measured conductance in the tunneling regime. Since the tunneling conductance is dominated by a molecular-scale volume at the point of closest interelectrode separation, these correlations imply few- or single-molecule Raman sensitivity. For this study, the conductances range from 0.1 to a few $G_0 = 2e^2/h$, and include contributions from both through-molecule tunneling and direct metal-metal tunneling or contact in some junctions. These junctions are not in the Coulomb blockade regime.

Figure 2 shows Stokes and anti-Stokes spectra as a function of applied bias for two representative devices. The main experimental observation is that many of the vibrational modes with energies greater than 1000 cm$^{-1}$ shift toward low energies as the applied bias increases. These systematic shifts are routinely observed in $C_{60}$-based junctions, having been seen in 12 out of 23 junctions that produced a significant and stable SERS signal. The remaining 11 junctions had "blinking" (junction configuration instability) sufficiently strong that it precluded the long measurements required for a clear assessment of bias-driven effects. This yield and variation is consistent with prior experiments in such junctions.

The bias-driven shifts, apparent as a curvature of the spectral features, vary in magnitude, from a few cm$^{-1}$ to 20 cm$^{-1}$. Figure 3 shows data from another device, employing a higher resolution grating in the spectrometer. This particular data set shows clear discontinuities in the mode intensities at a few bias voltages; these are stochastic blinking as described above. The bias-driven shifts on the devices are consistent with a quadratic dependence on applied bias, $\delta\omega \sim V^2$. Note that electromigrated junction experiments do not precisely control the molecule/metal contact geometry at the atomic scale; variability in the contact geometry and molecular environment can give junction-to-junction variations in the precise Raman spectrum. However, the sign, functional form, and magnitude of the bias-driven shifts here are consistent and reproducible.

To understand the mechanism at work, we use density functional theory (DFT) to compute the vibrational frequencies of $C_{60}$ as a function of external field and charge state. In principle, the local field and charge state of a molecule in a junction depend on atomistic features of the molecule-metal contact which are highly complex, generally unknown, and vary from device to device. Here, we neglect explicit treatment of the electrodes and instead model the $C_{60}$ environment, as a function of bias, through changes in fields and steady-state occupation. Initially, we compute the vibrational frequencies of a gas-phase $C_{60}$ molecule in the presence of constant electric fields. For several external fields up to 1.2 V/nm (approximately twice the range probed by the experiment), the $C_{60}$ geometry is relaxed and the vibrational modes are computed within DFT for constant charge state. Both neutral $C_{60}$ and the $C_{60}^{-1}$ anion are considered. Mirjani *et al.*(34) recently considered the impact of full reduction or oxidation on vibrational modes of molecules in junctions, observing appreciable shifts relative to the neutral species. However, the lack of resonant transport (see Supporting Information) confirms that the applied bias in this experiment is insufficient to fully change the average redox state of the molecule by an entire electron. Therefore, the anion represents the limit of charging possible in the system. For the neutral $C_{60}$, our calculations are in good agreement with previous theory and experiment (see SI). Calculated field-induced vibrational shifts for the neutral molecule (anion) are typically far less than 1 cm$^{-1}$ (5 cm$^{-1}$) in magnitude, at maximum field (1.2 V/nm); moreover, the shifts vary in sign and do not exhibit a generally quadratic functional form with field, in contrast with experiment. This rules out the well-known vibrational Stark effect as the origin of the observed phenomena.

A major clue toward an alternative explanation is that, at constant field, differences between specific mode frequencies of the neutral and anion are large, of order 10-150 cm$^{-1}$, and notably, the affected anion modes are red-shifted relative to the neutral molecule (see SI). This suggests an explanation in terms of bias-driven changes of the $C_{60}$ charge state. To explore this possibility, we recomputed the $C_{60}$ vibrational spectrum, adding small fractions of an electron, from 0 to 1, in steps of 0.1 $e$. In the case of significant hybridization between the molecule and metal contact, partially-occupied states in the junction are expected. Furthermore, in an open system a fractional number of electrons in DFT is defined via an ensemble of integer-electron states and interpreted as a time average of a fluctuating number of particles.(35) Therefore, one can infer the effect of partial molecular charging in the junction from calculations of a single partially-charged $C_{60}$ molecule. A fractional occupation of the $C_{60}$ LUMO upon adsorption into a junction is consistent with the established large electronegativity of $C_{60}$ (29, 36) and with STM studies of $C_{60}$ adsorbed on clean metal surfaces(37). As the molecule is (partially) charged, several $C_{60}$ vibrational modes shift systematically to lower energies, by tens of cm$^{-1}$. We compute that these are Raman-active $H_g$ modes[*] which couple strongly to the $t_{1u}$ LUMO(38-42) and are present throughout the 1000-1600 cm$^{-1}$ measurement range (see Fig. 4). This trend is reasonable on general chemistry grounds: Adding an electron to the neutral $C_{60}$ occupies an anti-bonding LUMO that is delocalized over the entire molecule, thereby softening many intramolecular bonds. Thus a redshift of vibrational modes coupled to an antibonding LUMO upon electronic charging would be expected quite generally.

---

[*] $A_g$ modes also couple to the $t_{1u}$ LUMO, but as the symmetric $A_g$ modes do not break the LUMO degeneracy, and are therefore not involved in the Jahn-Teller distortion(38-42), they vary less significantly with bias.

Only a relatively small amount of charging is necessary to result in the mode shifts seen here, and small changes in charge state are very plausible under bias. Figure 4 shows a simple model for the energy level alignment of the junction. At zero applied bias, the triply degenerate LUMO resonance will be positioned near the Fermi energy(37, 43), and broadened by its coupling to the source and drain electrodes. Assuming that electrons tunneling from the source to drain and drain to source, respectively; that they are non-interacting and are therefore occupied according to their original source or drain quasi Fermi levels(44); and that the resonance lineshape is Lorentzian with a width $\Gamma = \Gamma_S + \Gamma_D$, the change in steady-state occupation, $\delta\rho$, of a single triply degenerate level at energy $E_0$ above the equilibrium Fermi level $E_F$, at bias $V$, can be expressed as (1, 27)

$$\delta\rho = \int_{-\infty}^{\infty} \frac{1}{2} g(E) \left[ f\left(E + \frac{eV}{2}\right) + f\left(E - \frac{eV}{2}\right) - 2f(E) \right] dE \,, \tag{1}$$

where, in this case, the density of states $g(E) = (d\,\Gamma/\pi)/(\Gamma^2 + (E-E_0)^2)$ is Lorentzian with degeneracy $d$ ($d = 6$ in this case due to spin and orbital degeneracy) and $f(E) = 1/(e^{E/kT} + 1)$ is the Fermi-Dirac distribution function, with the zero-bias Fermi level $E_F$ taken as the energy reference, i.e. $E_F$=0. Here a symmetrical voltage drop is assumed, with the source and drain chemical potentials taken to be $\mu_S$=$eV/2$ and $\mu_D$=$-eV/2$, respectively. Recently, Kaasbjerg *et al.*(45) developed a non-equilibrium Green's function framework for understanding bias-dependent molecular vibrational mode damping and heating in junctions. Including physics similar to what we consider here with our $C_{60}$-specific model, this approach also captures vibrational frequency renormalization associated with charging and screening(45), and is consistent with the work presented here in the limit where $\Gamma$ is small compared to the resonance energy and bias voltage.

Previous STM experiments and DFT calculations of $C_{60}$ on metal surfaces yielded $\Gamma \sim 0.1$ eV and $E_0 \sim 1.0$ eV (43). However, in a junction environment, where $C_{60}$ is contacted on both sides with rough surfaces, $E_0$ will be closer to $E_F$ (37). Depending on the specifics of the Au-$C_{60}$ contact within a particular junction, $E_0$ may vary somewhat. The value for $\Gamma$ will also vary to some degree but numerous experiments have shown significant coupling of $C_{60}$ to Au. To demonstrate our reasoning, we take $E_0 = 0.6$ eV above $E_F$ at zero bias and $\Gamma = 0.1$ eV (the effect of other choices for these parameters is explored in Fig. S8 of the SI, but the general effect of charging is preserved). Together with the above model dependence of $\delta\rho$ on $V$ (Fig. 4b) and DFT-computed dependence of the frequency on $\delta\rho$ (Fig 4c), we compute the vibrational mode frequency as a function of bias for voltages up to +/-0.6 V, as shown in Fig. 4d. This simple model explains the measured mode softening trends.

For the choice of model parameters $E_0$ and $\Gamma$, the finite temperature spread of the Fermi-Dirac distribution of the electrons in the source and drain has a negligible effect on the molecular charge at 80 K, but is more important at 300 K. This suggests that any heating of the electronic distribution at high bias(23) could also play a role in determining the molecular charge and hence vibrational energies.

Using nanojunction-based SERS, we observe systematic bias-driven softening of vibrational modes in $C_{60}$. Comparisons with DFT calculations show that Stark physics alone cannot be responsible for these effects, and bias-driven alteration of the molecular charge state is the likely explanation. By combining realistic computational models of junctions with measurements of this type, the presence and degree of bias-induced mode softening can turn these junctions into a direct local probe of molecule/metal energetics. Interpreted in light of these observations, the earlier preliminary observations of bias-driven mode softening in a junction based on an oligophenylene vinylene (OPV) molecule(23) suggest that in that particular device the LUMO must lie close to the electrode Fermi levels. Indeed, recent calculations by Kaasbjerg et al. [41] arrive at similar conclusions regarding the origin of OPV mode shifts under bias observed in Ref. 5. Similarly, it is worth considering whether much of the spectral diffusion observed in single-molecule Raman measurements results from small changes in the effective molecular charge density, due to changes in the occupation of proximal surface states or the presence/absence of nearby molecular adsorbates. Finally, the observations reported here point out that considerable care should be taken in the interpretation of vibrational Stark effect data in other contexts. While good agreement between theoretical expectations and observations has been reported(11) when considering only Stark physics, it is important to note that charging effects can be of similar or greater magnitudes in some circumstances.

**Materials and Methods**

The SERS substrates for the measurements are produced by electromigration of lithographically defined metal constrictions fabricated on underlying oxidized Si substrates. The bowtie structures are patterned by electron beam lithography. The constriction of the bowtie structure is 100-120 nm wide and 1 micrometer long, as shown in the inset to Figure 1 of the main text. Evaporation of 1 nm Ti and 15 nm Au onto the bowtie structure is followed by lift off in acetone. The larger, extended contact pads for the bowtie structure are evaporated with 1 nm Ti and 30 nm gold using a shadow mask. The substrates are cleaned by oxygen plasma for 2 minutes to remove organic contamination and then spin coated with a solution of $C_{60}$ at a concentration of 1 mg of $C_{60}$ per 10 mL toluene. The $C_{60}$ molecules physisorb directly on the electrode surfaces, with no linker groups or linker chemistry. The chips are wire-bonded to a ceramic carrier and placed within a microscope cryostat. Once cooled to 80 K under a vacuum of 6 x $10^{-6}$ mB, each junction is electromigrated through a computer-controlled procedure to yield interelectrode tunneling conductances near or below $G_0 = 2e^2/h$.

Raman spectroscopy is performed using a home-built Raman microscope system, with an incident wavelength of 785 nm. Modes with Stokes or anti-Stokes shifts below 300 $cm^{-1}$ are cut off by a notch filter. A piezo-actuated lens mount rasters the diffraction-limited beam over the sample surface, allowing the acquisition of spatially mapped Raman response. The mapped Si Raman intensity at 520 $cm^{-1}$ is used to locate the center of a bowtie structure. Following electromigration, a further Raman image determines the location of the nanogap's plasmonic Raman hotspot. Raman spectra at that location are then acquired simultaneously with electronic transport data ($I$ and d$I$/d$V$ as a function of $V$, using a current preamplifier; $V$ sourced by a digital-to-analog converter (DAC) integrated into a lock-in amplifier; differential conductance measured via lock-in using a 10 mV AC signal added to $V$ with a summing amplifier). Fig. 1 is an example of a single surface enhanced Raman spectrum of such a junction, acquired with a 1 second integration time. The sharp mode at 520 cm$^{-1}$ is from the underlying Si substrate, and the modes between 1000 cm$^{-1}$ and 1600 cm$^{-1}$ agree reasonably well with expectations from other SERS studies of $C_{60}$. Transport data acquisition was synchronized with spectral measurements through triggering. The bias, $V$, was swept from -0.5 V to 0.5 V in steps of 0.0125 V or 0.025 V. The acquisition time for each spectrum at every voltage was 1-3 s. A higher resolution grating is available for detailed studies, though this grating precludes simultaneous measurements of both Stokes and anti-Stokes emission.

All calculations were performed using the Siesta 3.1 code(46). The C pseudopotential utilized a core radius cutoff of 1.29 Bohr for 2$s$, 2$p$, and 2$d$ channels. A triple zeta basis set was used for 2$s$ and 2$p$ functions, with a single zeta 2$d$ polarization function. All calculations were performed with a 1000 Rydberg real space grid and 30x30x30 Angstrom supercells. All structures at all charges and fields were relaxed until forces were less than 0.004 eV/Angstrom. Vibrations were calculated using SIESTA's Vibra package(47).

**Acknowledgments**: YL and DN acknowledge support from Robert A. Welch Foundation grant C-1636. Work by PD and JBN was supported by the U.S. Department of Energy, Office of Basic Energy Sciences, Materials Sciences and Engineering Division, under Contract No. DE-AC02-05CH11231. Portions of this work at the Molecular Foundry were supported by the Office of Science, Office of Basic Energy Sciences, of the U.S. Department of Energy under the same contract number. Computational resources provided by NERSC. Work by LK was supported by the Israel Science Foundation and the Lise Meitner Center for Computational Chemistry.

**Author Contributions**:  YL fabricated the samples, performed the experiments, and analyzed the data.  DN conceived of the experiments and guided the analysis.  PD performed the modeling calculations, in consultation with JN and LK.  All authors contributed to the manuscript.

*Figures*:

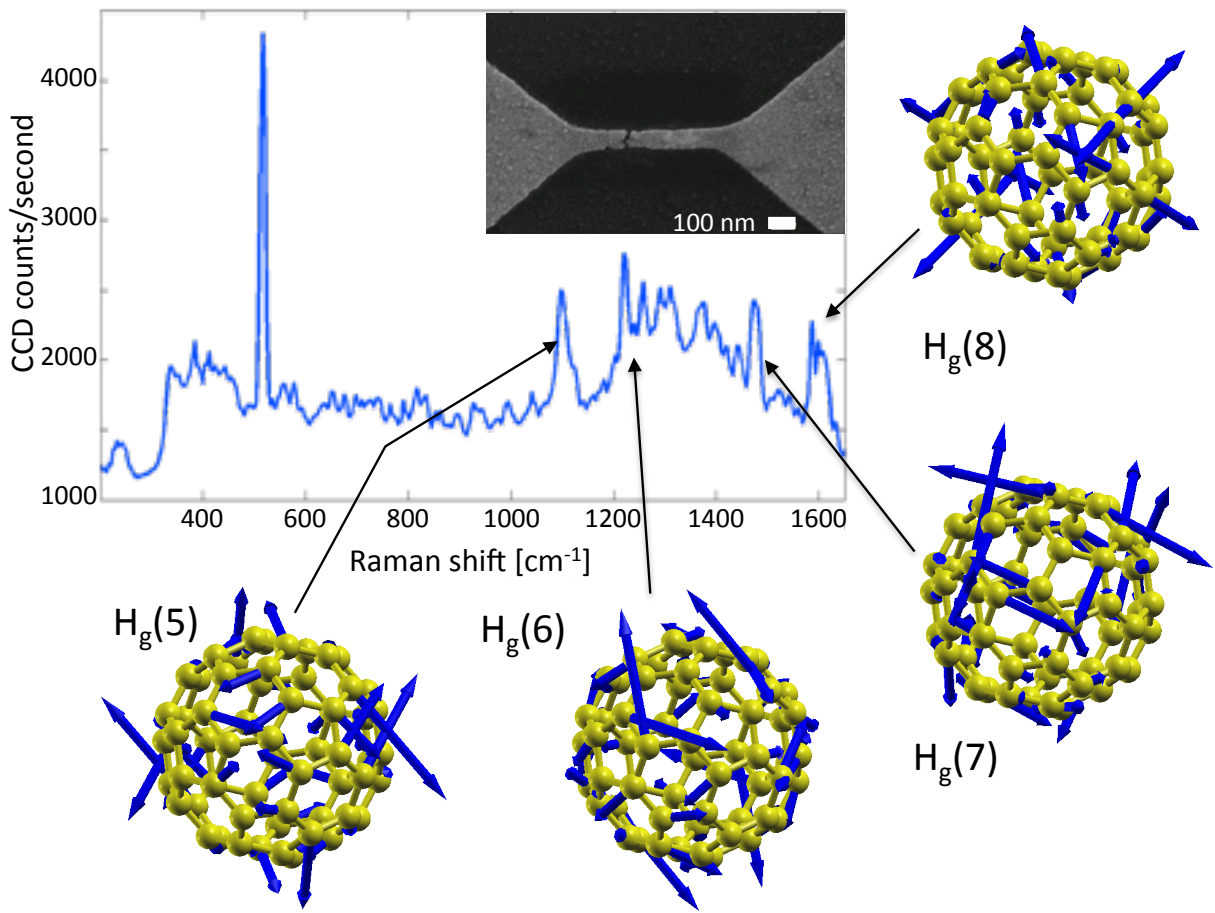

**Figure 1**: **Raman analysis of C$_{60}$ in an electromigrated junction**. Main plot: Example SERS spectrum of C$_{60}$ in an electromigrated junction. Surrounding diagrams illustrate examples of the complicated displacements associated with Raman-active modes, calculated for an isolated, symmetric, gas phase molecule. Each such mode is five-fold degenerate in the absence of symmetry breaking. Inset: Scanning electron microscopy image of an electromigrated junction.

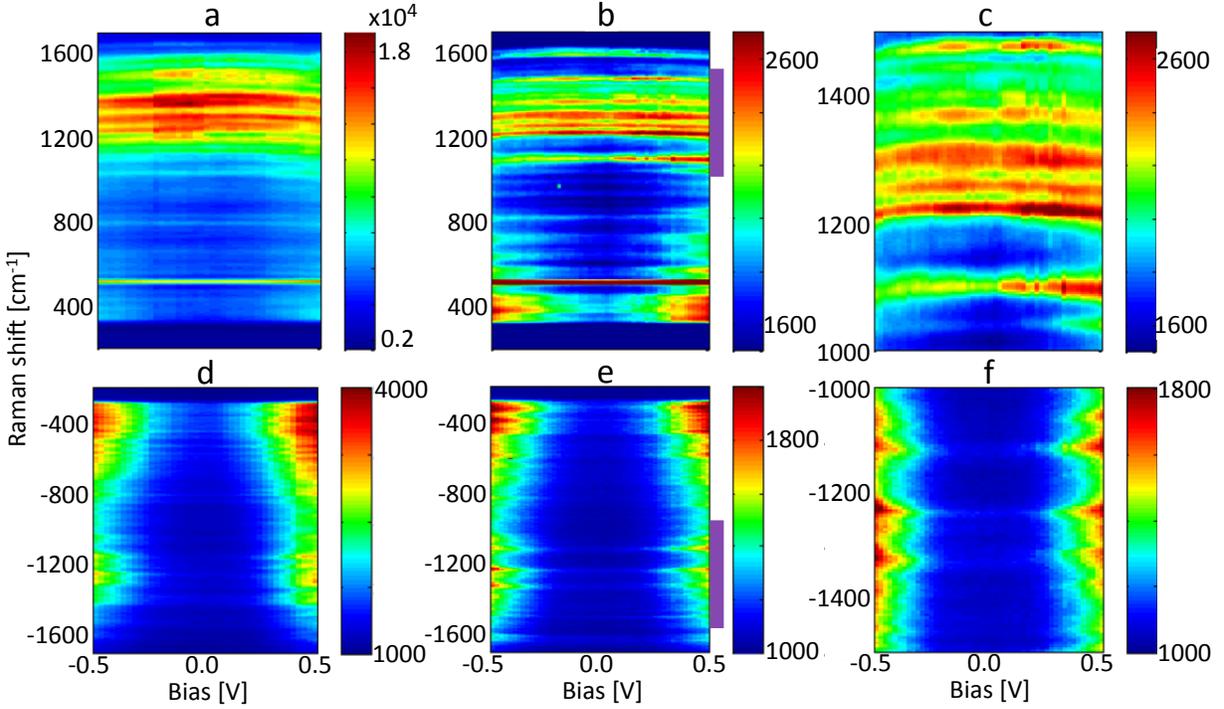

**Figure 2**: **Vibrational modes and their evolution with source-drain bias**. (a) Stokes and (d) anti-Stokes spectra of Device 1 as a function of $V$. (b) Stokes and (e) anti-Stokes spectra of Device 2. Color scales indicate counts per integration time. (c) and (f) are rescaled closeups of the Device 2 data over the wavenumber ranges indicated by the purple bars in (b) and (e), respectively. The vibrational modes curve slightly toward to lower energies at larger $|V|$, and the anti-Stokes intensities increase at high biases. The latter effect indicates current-driven heating of vibrational degrees of freedom, as reported previously(22, 23). The spectral shifts are more difficult to resolve in the anti-Stokes case because of this evolution of anti-Stokes intensity.

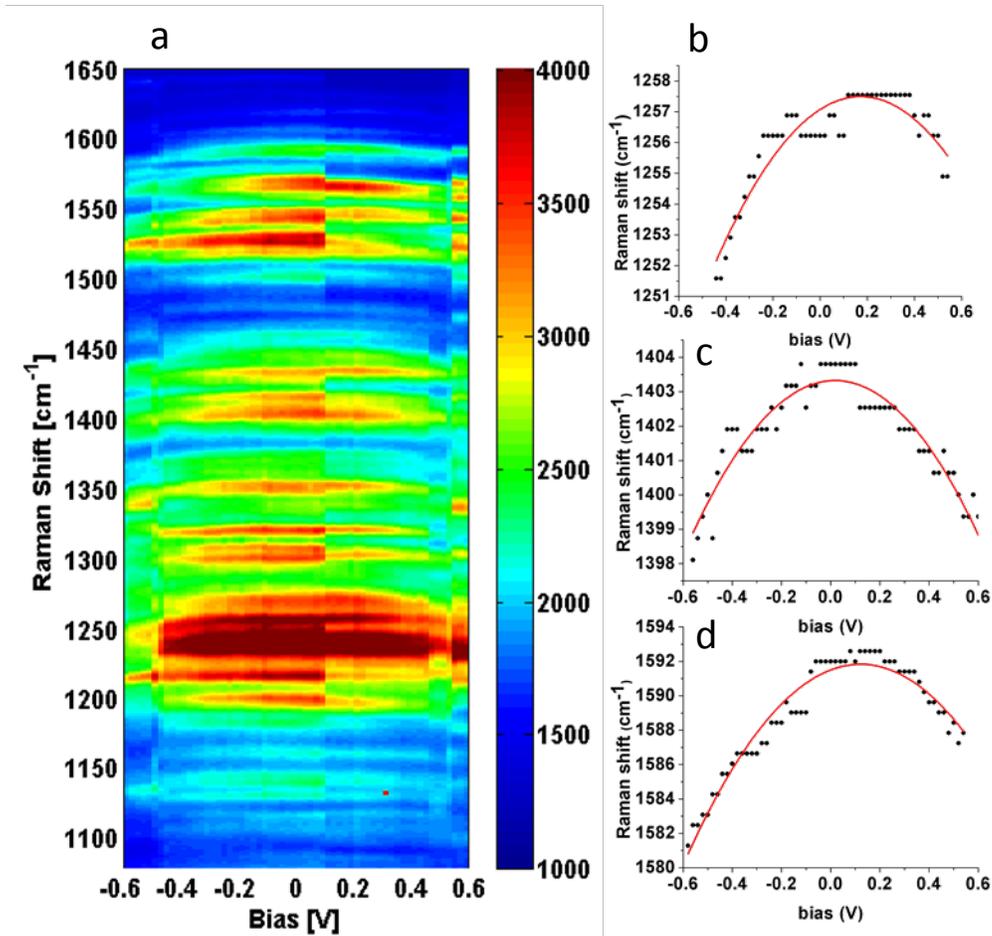

**Figure 3**: **Bias driven vibrational energy shifts**. (a) Raman response of Device 3 as a function of bias (x axis) and Raman shift (y axis). The sudden change of the intensity at around 0.1 V is due to blinking. (b, c, d) vibrational energy shift as a function of bias for three particular modes: 1258 cm⁻¹, 1404 cm⁻¹ and 1592 cm⁻¹. The discretized Raman shift data results from pixilation of the detector.

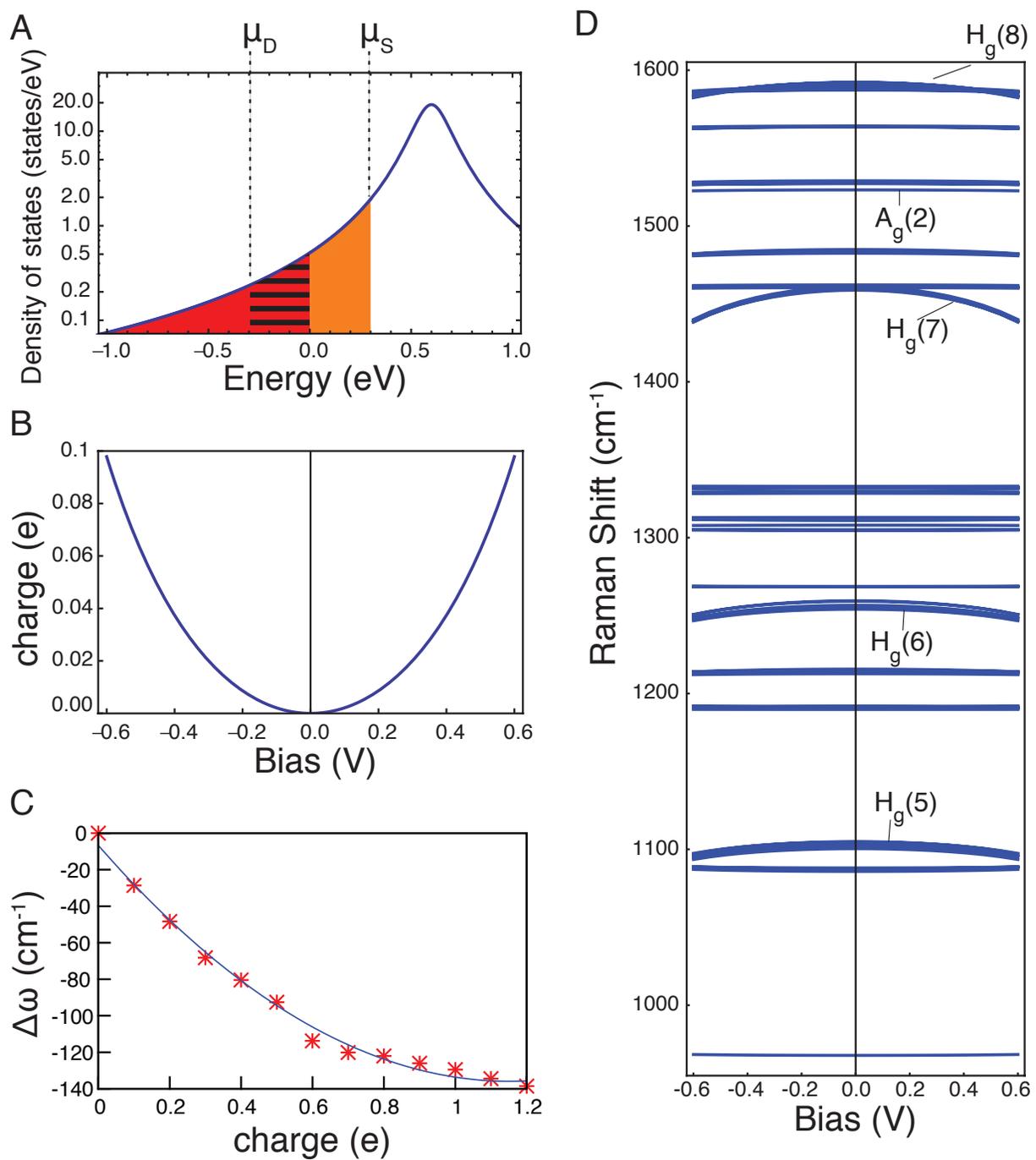

**Figure 4**: **Model of bias-driven changes in molecular charging**. (a) At zero bias, the triply degenerate LUMO resonance, centered at $E_0$ with width $\Gamma$, is occupied proportionally to the red shading. As the bias $V$ is applied, the molecular level gains additional occupation proportional to the area shown by the orange shading and loses occupation proportional to the hatched portion of the Lorenztian. (b) The expression for charging with bias at 80 K (solid) is visually identical to the charging at 0 K (for 300K charging see SM). The change in partial charge is approximately quadratic in bias. (c) A representative mode's (Hg(7) at 1467 cm$^{-1}$) change in vibrational energy with charging, computed via DFT. This dependence, combined with the variation in charge with bias, strongly suggests that bias-driven charging is the origin of the systematic mode softening observed in the experiments. (d) Mode energies as a function of bias from such a calculation.

**Supporting Information for "Voltage tuning of vibrational mode energies in single-molecule junctions"**

Yajing Li[1], Peter Doak[2,3], Leeor Kronik[4], Jeffrey B. Neaton[3,5,6], Douglas Natelson[1,7*]

[1]*Department of Physics and Astronomy, MS 61, Rice University, Houston, TX 77005*

[2]*Department of Chemistry, University of California at Berkeley, Berkeley, CA 94720*

[3]*Molecular Foundry, Lawrence Berkeley National Laboratory, Berkeley, CA 94720*

[4]*Department of Materials and Interfaces, Weizmann Institute of Science, Rehovoth 76100, Israel*

[5]*Department of Physics, University of California at Berkeley, Berkeley, CA 94720*

[6]*Kavli Energy Nanosciences Institute at Berkeley, Berkeley, CA 94720*

[7]*Department of Electrical and Computer Engineering, MS 366, Rice University, Houston, TX 77005*

*natelson@rice.edu.

**Current-Voltage Characteristics**

Bias-driven changes in the molecular charge can explain the systematic mode softening reported in the main manuscript. This explanation assumes that the lowest unoccupied molecular orbital (LUMO) lies sufficiently above the (unbiased) Fermi level of the source and drain electrodes that the applied bias is insufficient to bring the level fully into the bias window. Therefore, one would expect the current-voltage characteristics of the junctions to be nonlinear, with increased conductance at high bias, but only weakly so, since the LUMO remains outside the bias window even at the maximum applied voltage. Figures S1, S2, and S3 are measured *I-V* characteristics, averaged over multiple bias sweeps (to minimize the effects of the stochastic "blinking" due to variation in the atomic-scale details of the junction). This amount of variation in *I-V* characteristics is typical of what is seen in electromigrated molecular junctions. The devices are not in the Coulomb blockade regime and there are no signs of resonant conduction. Furthermore, as stated in the main text, the conductance in general includes contributions from both through-molecule tunneling and direct metal-metal transport. This is different than the situation in mechanical break junction experiments, which deliberately avoid metal-metal contributions.

**Computational Details**

Table S1 shows a comparison against previous theoretical results for the vibrational spectrum of $C_{60}$ for Raman active modes. Some difference between the results from the previous calculations done with the Gaussian code and this work remain even at the high degree of convergence our parameters detailed above achieve. We ascribe this primarily to Siesta's differing construction of the basis through the use numerical atomic orbitals (NAO). Importantly, the difference in PBE results between the two codes is much smaller than the difference between PBE and B3LYP results. This indicates that, as required, the results are indeed dominated by the choice of exchange-correlation rather than the details of the numerical simulation.

For calculations of charged $C_{60}$ Siesta's built in mechanism for charged systems was utilized. It uses a compensating uniform background charge and a Madelung correction. For the finite field calculations the fields were applied along $C_2$ symmetry axis and the $C_{60}$ was allowed to relax at each field strength. Modes were tracked across the charge state calculations of the molecule by tracking the strongest projection for each mode as charge was increased.

In Fig. S5 we show the result of varying $E_0$ on the simulated vibrational spectrum as bias is swept.

**Table S1.**

Raman Active Neutral C60 Vibrational Modes in $cm^{-1}$ (percent error vs. experiment). For Hg modes value is average of 5 nearly (+/- 1 $cm^{-1}$) degenerate modes.  * Previous results calculated in the Gaussian code with 6-311g(d) basis set (J. L. Janssen, M. Côté, S. G. Louie, and M. L. Cohen, Electron-phonon coupling in $C_{60}$ using hybrid functionals. *Phys. Rev. B.* **81,** 073106 (2010)).

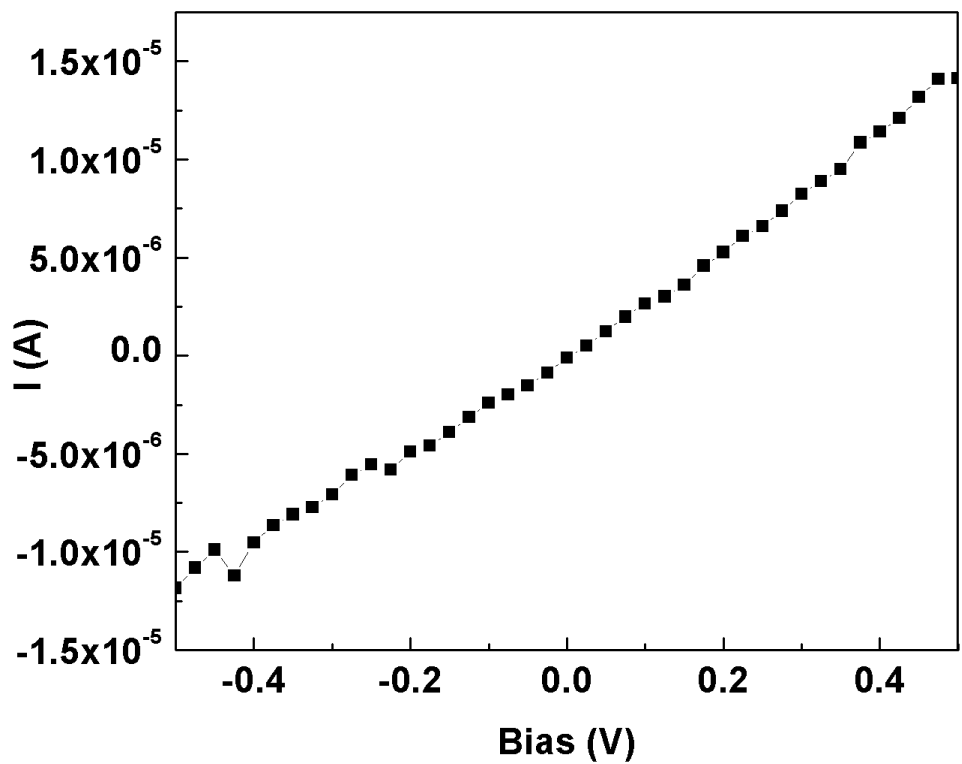

**Figure S1.** Current-voltage characteristic for the junction corresponding with the Raman spectra presented in Figs.

2a and d of the main manuscript.

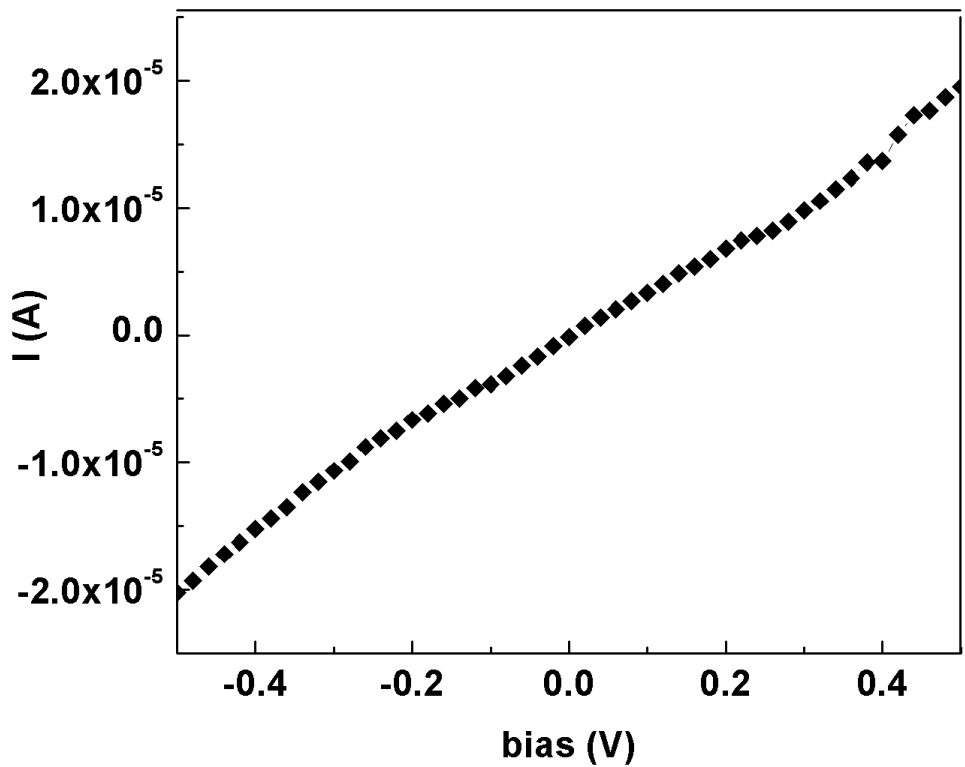

**Figure S2.** Current-voltage characteristic for the junction corresponding to the Raman data shown in Figs. 2b, c, e, f of the main manuscript.

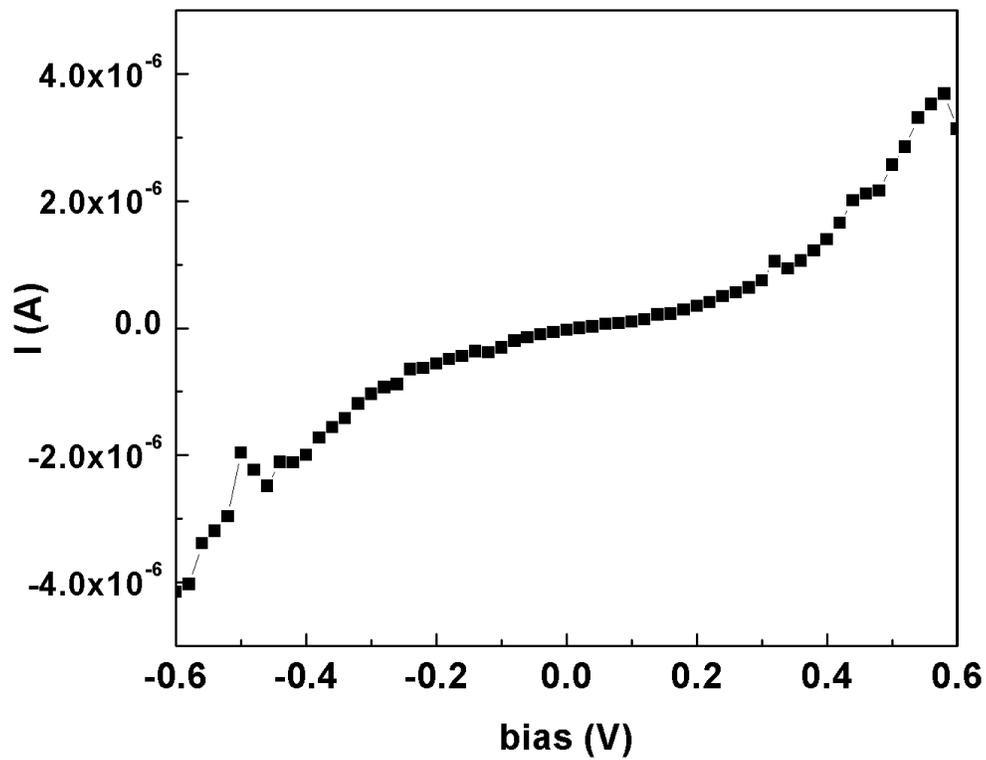

**Figure S3.** Current-voltage characteristic for the junction with Raman spectra shown in Fig. 3 and Fig. 4 of the main manuscript.

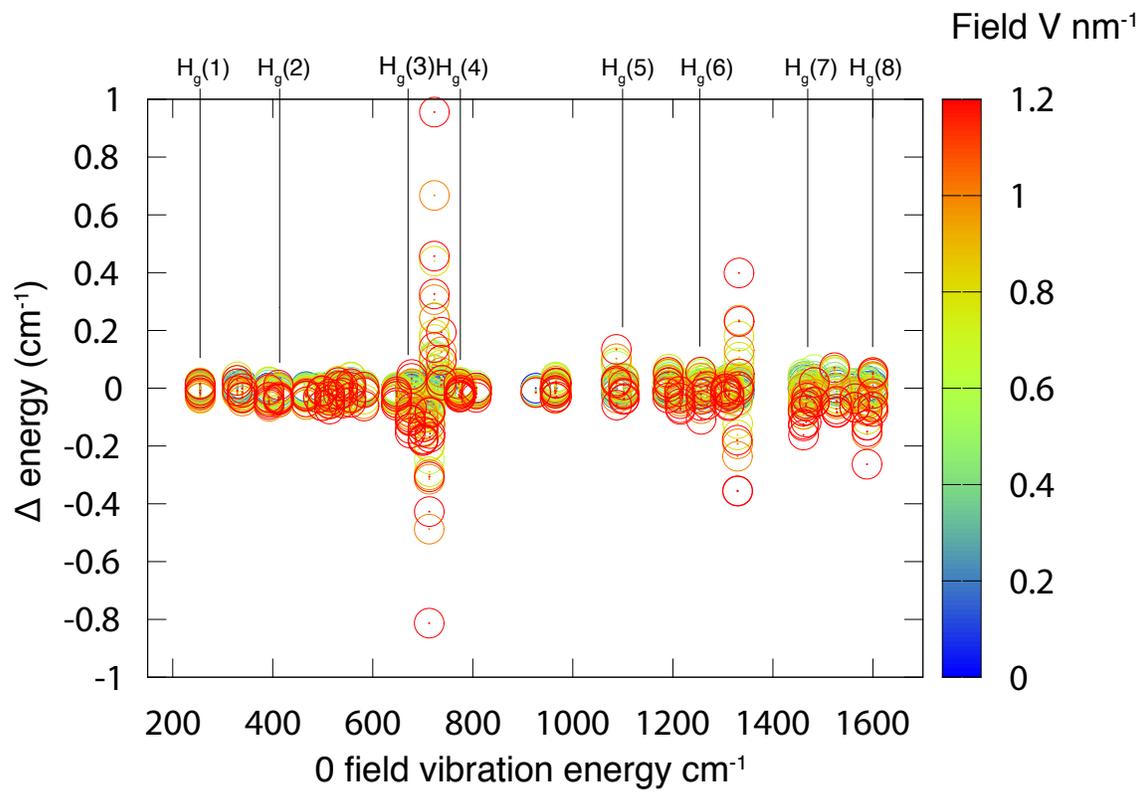

**Figure S4.** Shift in vibrational energy for neutral $C_{60}$ with electric field.

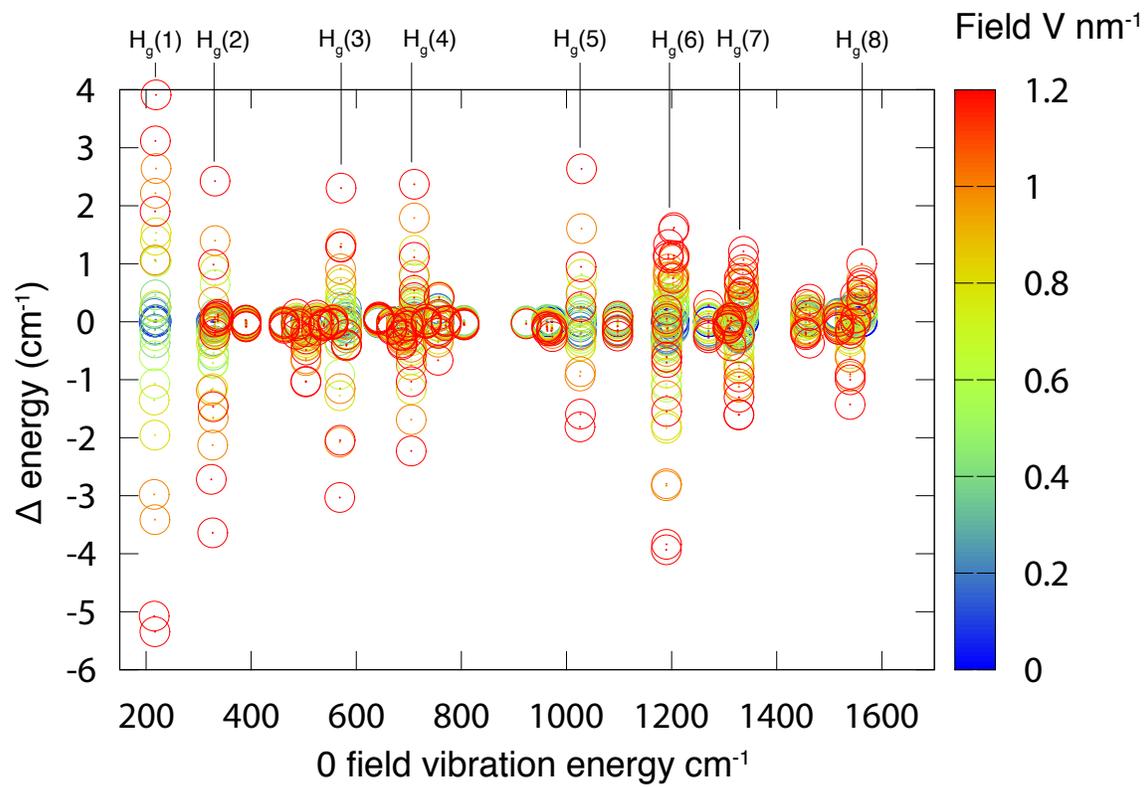

**Figure S5.** Shift in vibrational energy for $C_{60}$ anion with electric field.

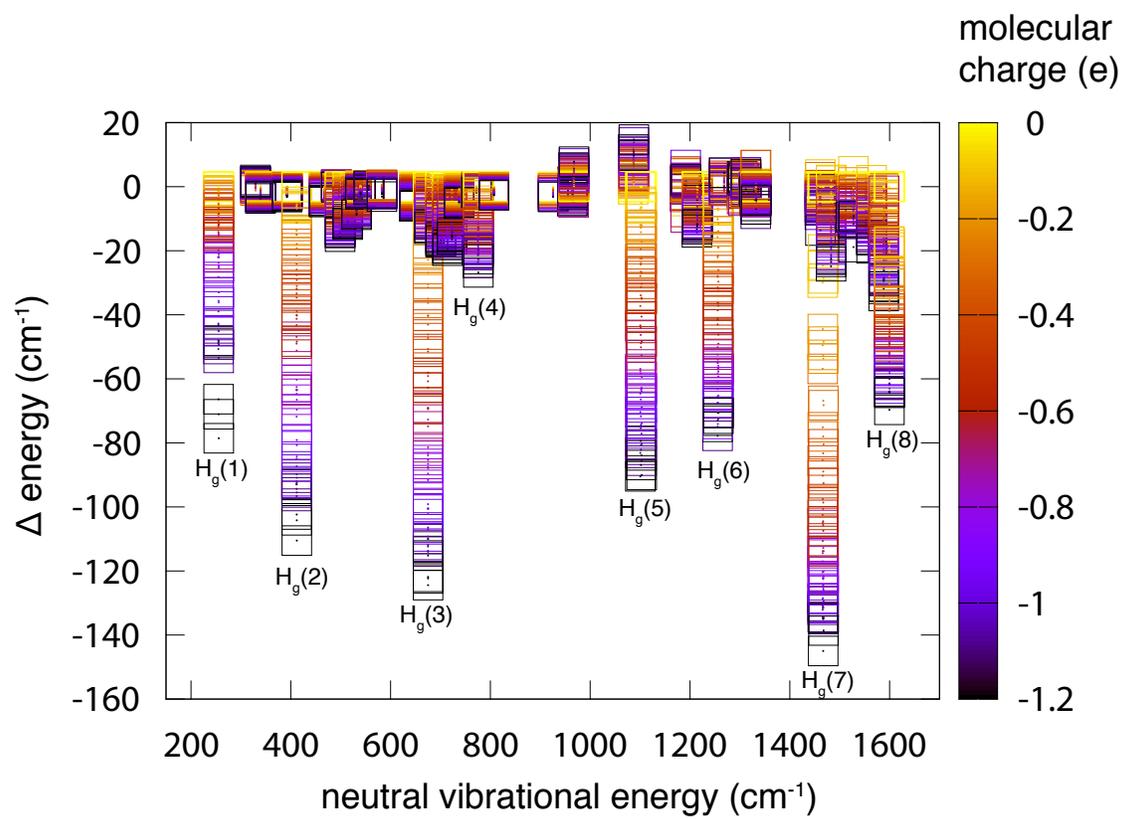

**Figure S6.** Shift in vibrational modes with partial charging.

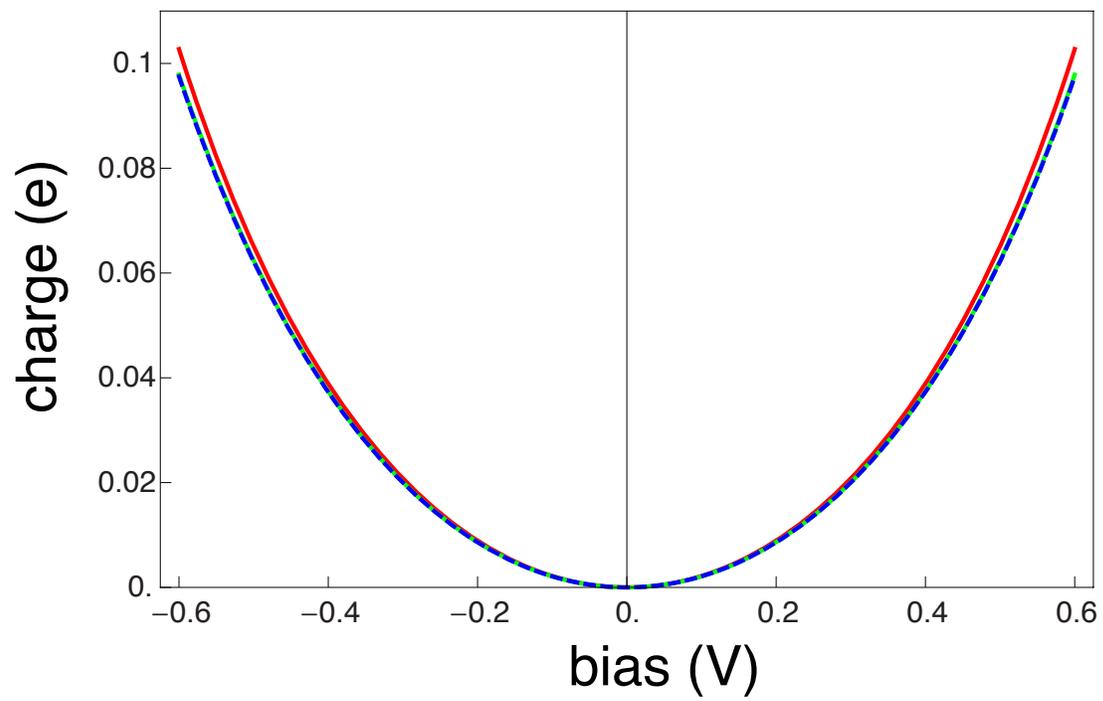

**Figure S7.** Charging of C$_{60}$ with single Lorentzian model at 300K (red), 80K (green), and 0K (dashed blue).

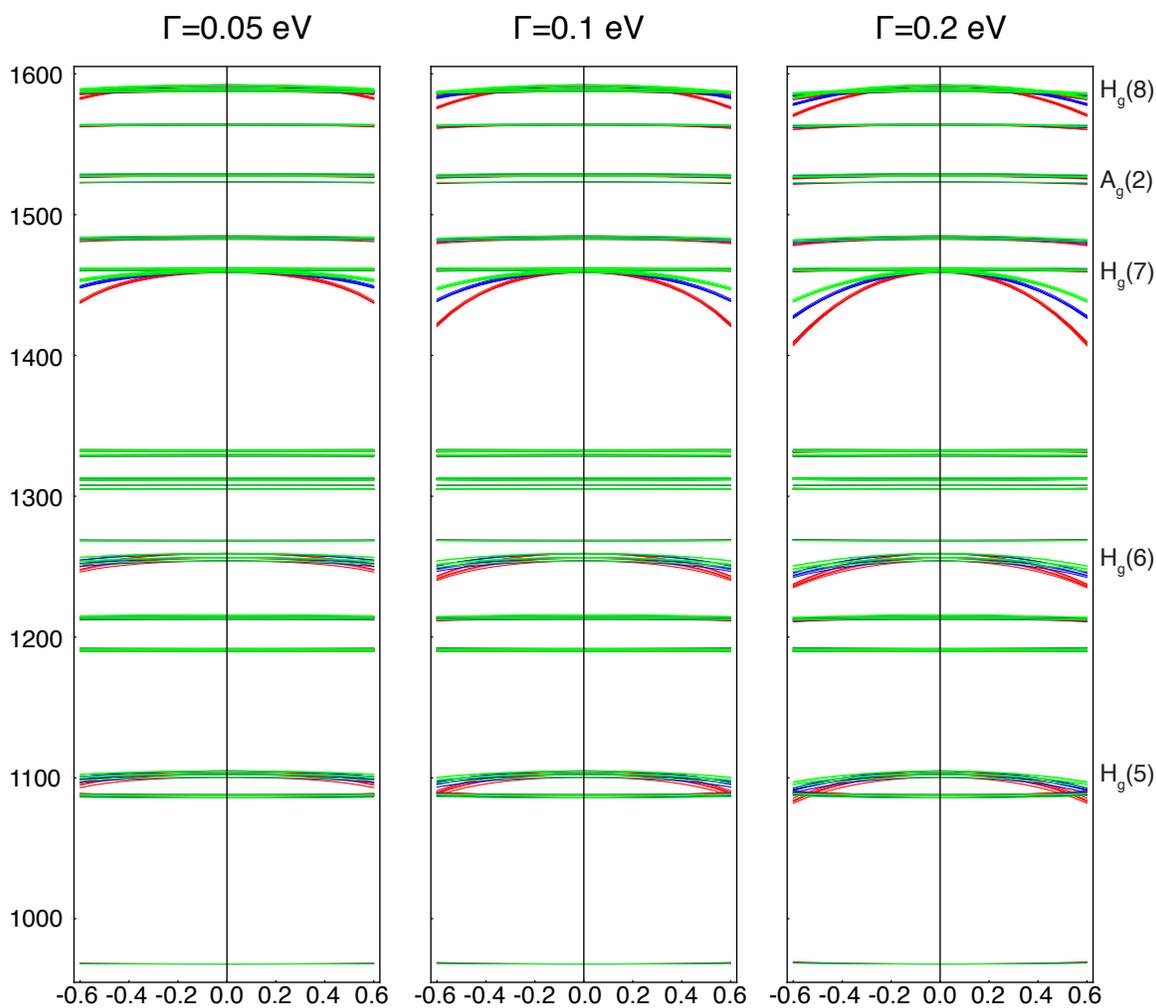

**Figure S8.** Modeled spectrum at $E_0$=0.5 eV (red), 0.6 eV (blue), 0.7 eV (green) and $\Gamma$=0.05 eV, 0.1 eV, 0.2 eV